\documentclass[aps,prb,twocolumn,showpacs]{revtex4}
\usepackage[utf8]{inputenc}
\usepackage[english]{babel}
\usepackage{xcolor}
\usepackage{braket}
\usepackage{bm,amsmath}
\usepackage{amssymb}
\usepackage{graphicx}
\newcommand{\pda}{^{\phantom\dagger}}
\newcommand{\da}{^{\dagger}}
\renewcommand{\vec}{\textbf}

\begin{document}
     \title{Quantum dot exciton dephasing by Coulomb interaction. A 
fermionic analogue of the independent boson model}

    \author{I.V. Dinu$^{1,2}$, M. \c Tolea$^1$, and P. Gartner$^2$}

    \affiliation{$^1$ National Institute of Materials Physics, Atomistilor 
405A, M\u agurele 077125, Romania}
    \affiliation{$^2$ Centre International de Formation et de Recherche 
Avanc\'ees en Physique - NIMP, Atomistilor 407, M\u{a}gurele 077125, Romania}
    \date{\today}

    \begin{abstract}
The time evolution of a quantum dot exciton in Coulomb interaction with wetting 
layer carriers is treated using an approach similar to the independent boson 
model. The role of the polaronic unitary transform is played by the scattering 
matrix, for which a diagrammatic, linked cluster expansion is available. 
Similarities and differences to the independent boson model are discussed. A 
numerical example is presented.
    \end{abstract}  
      
  \maketitle    
  
\section{Introduction}
Quantum dots (QD) in semiconductor heterostructures are sometimes regarded as 
artificial atoms, but one aspect in which they differ from real atoms is that 
they live in an invasive environment. QD carriers interact with both acoustic 
and optical phonons, and sometimes with free carriers in the wetting layer (WL) 
or in the bulk. Such interactions play an important role in the QD optical 
properties, and especially in the dissipative behavior, like carrier population
redistribution and polarization dephasing. 

In this respect, the role of the phonons enjoyed a much larger attention in the 
literature, one reason being the availability of the independent boson model 
(IBM) \cite{mahan,huang_rhys, duke_mahan}, which is both simple and in 
certain circumstances exact. The 
popularity of the method cannot be overestimated, it being used in various 
contexts like the theory of exciton dephasing and absorption \cite{zimm_ebg, 
wacker,li_jqe}, phonon-assisted exciton recombination \cite{hofling}, 
phonon-mediated off-resonant light-matter coupling in QD lasers 
\cite{florian_1, hohenester}, generation of entangled phonon states 
\cite{hahn_kuhn} or phonon-assisted adsorption in graphene \cite{sengupta}, to 
cite only a few.  

The IBM relies on the carrier-phonon interaction being diagonal in the 
carrier states. If not, or if additional interactions are present, the method 
ceases to be exact, but it is still helpful as a way to handle a part of the 
interaction, responsible for the polaron formation. 
This diagonality implies that the bosons see an occupation number, either zero 
or one, according to whether the QD excitation is present or not. The 
difference 
between the two cases amounts, in a linear coupling, to a displacement of the 
phonon oscillation centers, without changing their frequency. In other words 
one has just a change of basis, performed by a unitary operator, the polaronic 
transform. 
The diagonality requirement means that the method is particularly suited for 
calculating pure dephasing, i.e. polarization decay without population changes, 
and absorption spectra \cite{zimm_ebg, besombes}, where it produces exact 
analytic results.

In view of this wide range of applications it is legitimate to ask whether such 
a unitary transform does not exist for the interaction with the free carriers 
too. 
We show that the answer is positive, if we frame the problem as above, i.e. if 
the interaction is diagonal in the QD states. We illustrate the situation in 
the case of a two-level system where the electron-hole vacuum acts as the 
ground state and an exciton pair as the excited state. The charge distribution 
of the latter acts as a scattering center for the carriers in the continuum. It 
is known \cite{taylor, sitenko} that the free and the scattered continua are 
unitary equivalent, with the transform provided by the scattering matrix. In 
many ways this approach resembles the IBM, and can be regarded as its fermionic 
analogue. Many similarities between the two can be seen, but we also point out 
the 
differences. Most importantly, the method is not exact any more, but it lends 
itself to a diagrammatic expansion. 

In a previous paper \cite{florian_2} this approach has been already used in the 
context of a nonresonant Jaynes-Cummings (JC) model. The cavity feeding was 
assisted by the fermionic bath of WL carriers, which compensated the energy 
mismatch. The situation there is more complicated due to the QD states getting 
mixed by the JC interaction with the photonic degrees of freedom.  

In the present paper, we address a simpler, more clearcut situation, involving 
only the exciton-continuum interaction. We illustrate the method by calculating 
the polarization decay and absorption line shape, as functions of bath 
temperature and carrier concentration. We also discuss in more detail the 
similarities and differences with respect to the IBM.

\section{The model}
\label{sec:model}

The system under consideration consists of a QD exciton interacting with a 
fermionic thermal bath, represented by the WL carriers. The latter are taken 
interactionless and in thermal equilibrium.
The Hamiltonian describing the problem is ($\hbar=1$)
\begin{align}
\begin{split}
    H = &\,\varepsilon\pda_X X^\dagger X + (1 - X^\dagger X)\,h_0 + X^\dagger X 
\, h_X\,, \\ 
  h_0 = & \sum_{\lambda=e,h}\sum_{\vec k} \varepsilon^{\lambda}_{\vec k}
        \lambda^\dagger_{\vec k} \lambda\pda_{\vec k}\,,  \quad\quad h_X = h_0 
+ W \, .
            \label{eq:hamiltonian}
\end{split}            
\end{align}
Here $X\da$, $X$ are the excitonic creation and annihilation operators. 
Specifically, considering in the QD one $s$-state in each band, with operators 
$e\da,e$ and $h\da,h$ for electrons and holes respectively, we have $X=he$. 
Limiting the QD configurations to the neutral ones, one has $X\da X=e\da e=h\da 
h$.  
The exciton energy is $\varepsilon\pda_X$. The WL Hamiltonian is given by 
$h_0$, with the subscripted $\lambda$ symbol meaning either $e$ or $h$ 
WL-continuum operators, and the momentum index $\vec k$ including tacitly the 
spin. In the presence of the exciton the WL Hamiltonian $h_X$ gets additionally 
a term $W$, describing the interaction with the QD carriers.    

This is expressed in terms of the matrix elements 
\begin{equation}
V^{\lambda\lambda'}_{ij,kl} = \sum_{\vec q} V_{\vec 
q}\bra{\varphi^{\lambda}_i} 
 e^{i\vec q\vec r}\ket{\varphi^{\lambda}_l}\bra{\varphi^{\lambda'}_j} e^{-i\vec 
q\vec r}\ket{\varphi^{\lambda'}_k}
 \label{eq:mat_el}
\end{equation}
of the Coulomb potential $V_{\vec q}$, between one-particle states, with 
$\lambda, \lambda'=e,h$. Only two of these indices are needed 
since band index conservation is, as usual, assumed.

The WL-QD interaction has the form
\begin{align}
\begin{split}
 & X\da X \,W=  \\ 
 & \sum_{\vec k, \vec k'} e\da_{\vec k} e\pda_{\vec k'} 
                  \left[ V^{ee}_{s \vec k, \vec k's} e\da e 
                        -V^{he}_{s \vec k, \vec k's} h\da h
                        -V^{ee}_{s \vec k, s \vec k'} e\da e \right] + \\  
 & \sum_{\vec k, \vec k'} h\da_{\vec k} h\pda_{\vec k'} 
                  \left[ V^{hh}_{s \vec k, \vec k's} h\da h 
                        -V^{eh}_{s \vec k, \vec k's} e\da e
                        -V^{hh}_{s \vec k, s \vec k'} h\da h \right] \, . 
                        \label{eq:W}
\end{split}
\end{align}
It shows that the WL carriers are scattered by an external field produced by 
the exciton, having the form
 \begin{equation}
 W = \sum_{\lambda=e,h}\sum_{\vec k, \vec k'} W^\lambda_{\vec k, \vec k'} 
\lambda \da_{\vec k} \lambda \pda_{\vec k'} \,.
 \label{eq:scat_field}
 \end{equation}
 In each $W^\lambda_{\vec k, \vec k'}$ the first two terms describe direct, 
electrostatic interaction between WL-carriers with the QD electron and hole 
respectively. The difference between repulsion and attraction is nonzero due to 
the different charge densities of these two. The exciton is globally neutral, 
but local charges are usually present and generate scattering. The strength 
of the scattering depends on the degree of charge compensation within the 
exciton. 
The third term is the exchange contribution, and it is not expected to be 
large, 
since around $\vec q =0$ the matrix elements in Eq.\eqref{eq:mat_el} become 
overlap 
integrals between orthogonal states.     
    
The idea of the present method relies on the unitary equivalence between the WL 
Hamiltonians $h_0$ and $h_X$ provided by the scattering matrix $\mathcal 
S(0,-\infty)$ (see e.g. ~\cite{taylor,sitenko}). One has
    \begin{equation}
        \mathcal S(-\infty,0) \, h_X \, \mathcal S(0,-\infty) = h_0 \, ,
        \label{eq:hh0}
    \end{equation}
    with $\mathcal S$ generated by the scattering potential $W$
    \begin{align}
        \begin{split}
            \mathcal S(t_1, t_2) &= \mathcal S^\dagger(t_2,t_1)\\
        &=\mathcal{T} \exp \left[ -i \int_{t_2}^{t_1} \, W(t)\,\text{d}t 
\right] \,, \quad t_1>t_2 \, .
        \end{split}
    \end{align}
$\mathcal{T}$ is the time ordering operator and the interaction representation 
of the perturbation $W(t)$ with respect to $h_0$ is used. 

For the full WL-QD Hamiltonian we formally eliminate the interaction part using 
the unitary transform
\begin{equation}
U = 1-X^\dagger X + X^\dagger X \,\mathcal S(0,-\infty) \, ,
\end{equation}
which switches on the action of the $\mathcal S$-matrix only when the exciton 
is present. It follows that
   \begin{align}
        \begin{split}
            U^\dagger & \left[(1-X^\dagger X)\, h_0 + X^\dagger X \, h_X 
\right] U = \\
            (1-X^\dagger X)\, h_0  
            + & X^\dagger X\, \mathcal S(-\infty,0)\,h_X\, 
            \mathcal S(0,-\infty) = h_0 \, .
        \end{split}
    \end{align}    
The excitonic operators are changed, according to $\widetilde X= U\da\, X\, 
U=X\, \mathcal S(0,-\infty)=\mathcal S(0,-\infty)\, X$ and similarly 
$\widetilde X\da= X\da \mathcal S(-\infty,0) $,  but $X\da X$ remains 
invariant. Therefore in the transformed problem
\begin{equation}
\widetilde H= U\da\, H\, U = \varepsilon\pda_X X\da\, X + h_0\, ,
\label{eq:H_tilde}
\end{equation}
the QD and the WL become uncoupled. 
This follows faithfully the effect produced by the polaronic transform in the 
bosonic bath case, as described by the IBM.
    
Assuming an instantaneous excitation at $t=0$, the linear optical polarization 
of the QD is expressed in terms of the exciton retarded Green's function    
\begin{equation}
\mathcal P(t)= -i\, \theta(t) \braket{X(t) X\da} = 
-i\,\theta(t)\text{Tr}\{\rho\, X(t) X\da \}\, .
\end{equation}
In the unitary transformed picture 
\begin{equation}
\mathcal P(t)= -i\, \theta(t)\text{Tr}\{\widetilde \rho\, \widetilde X(t) 
\widetilde X\da \} \, ,
\end{equation}
the problem is interaction-free, and QD and WL operators evolve independently. 
Therefore
\begin{align}
\widetilde X(t) & = e^{-i \varepsilon\pda_X t} X e^{i h_0 t} \mathcal S(0, 
-\infty) e^{-i h_0 t} \nonumber \\
&= e^{-i \varepsilon\pda_X t} X 
       \mathcal S (t,-\infty)\,,
\end{align} 
and as a consequence
\begin{equation}
\mathcal P(t)= -i\, \theta(t) e^{-i \varepsilon\pda_Xt} 
\braket{XX\da}\braket{\mathcal S(t,0)} \, .
\end{equation}
Essentially, besides a trivial exciton energy oscillation, the problem boils 
down to evaluating the thermal bath average of the scattering matrix $\mathcal 
S(t,0)$ for positive times.  

\begin{figure}[t]
        \centering
        \includegraphics[width=\columnwidth]{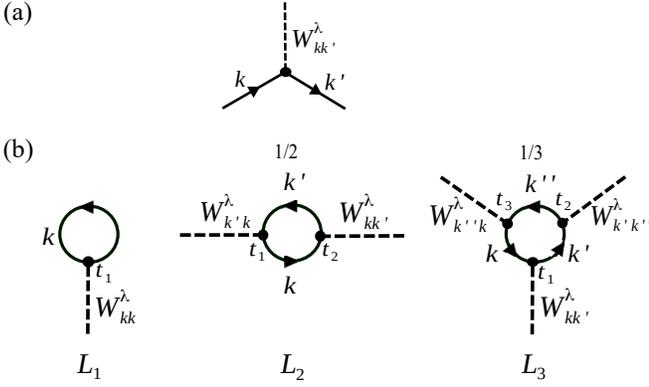}
        \caption{(a) Elementary interaction vertex. (b) First three connected 
diagrams $L_1$, $L_2$ and $L_3$ of the linked cluster expansion in the 
evaluation of the thermal average of the $S$-matrix $\braket{S(t,0)}$.}
        \label{fig:diagrams}
\end{figure}

This is again formally similar to the IBM case, but differing in the details, 
as will be discussed below. In the present case one makes use of the linked 
cluster (cumulant) expansion for $\braket{\mathcal S (t,0)}$ 
\cite{abrikosov,mahan}, in which a lot of resummation has been 
performed. As a consequence $\braket{\mathcal S(t,0)}$ is expressed  
as an exponential, $\exp[\Phi(t)]$, where $\Phi(t)$ is the sum of all connected 
diagrams with no external points $\Phi(t)=\sum_n L_n(t)$, where the diagram 
$L_n\, , n=1,2,3 \dots$  of order $n$ comes with a factor $1/n$. 
Its internal points are time-integrated from $0$ to $t$. In our case the 
interaction is an external potential, not a many-body one and the elementary 
interaction vertex in the diagrams is as in Fig.~\ref{fig:diagrams}(a). The 
first diagrams of the expansion are represented in Fig.~\ref{fig:diagrams}(b). 
One has

\begin{align}
        \begin{split}
            L_1(t) =& - \sum_{\lambda,\vec k} \int_0^t \text{d}t_1 
W^{\lambda}_{\vec k,\vec k}\, G^0_{\lambda \vec k}(t_1,t_1^+) \\
            L_2 (t)=& -\frac{1}{2} \sum_{\lambda, \vec k,\vec k'} 
\Big|W^{\lambda}_{\vec k,\vec k'}\Big|^2 \\
            &\times\int_0^t \text{d}t_1\int_0^t\text{d}t_2 \,G^0_{\lambda \vec 
k}(t_1,t_2) \, G^0_{\lambda \vec k'}(t_2,t_1)\,,
        \end{split}
        \label{eq:diag_uneval}
    \end{align}
    where $G^0_{\lambda \vec k}$ is the free Green's function for the WL state 
$\lambda \vec k$. 

For the first-order contribution one obtains an imaginary, linear time 
dependence, which amount to a correction to the exciton energy.

\begin{equation}
        L_1(t) = - i\sum_{\lambda,\vec k} W^{\lambda}_{\vec k,\vec k}\, 
f^\lambda_{\vec k}\, \,t
\label{eq:L1}
\end{equation}
with $f^\lambda_{\vec k}$ the Fermi function for the WL level carrying the 
same indices.

More important is the second order diagram ($\varepsilon^\lambda_{\vec k 
\vec k'}$ denotes the difference $\varepsilon^\lambda_{\vec 
k}-\varepsilon^\lambda_{\vec k'}$)

 \begin{align}
 \begin{split}
        L_2(t) =&\sum_{\lambda,\vec k,\vec k'} \Big| W^{\lambda}_{\vec 
k,\vec k'}\Big|^2 (1-f^{\lambda}_{\vec k})f^{\lambda}_{\vec k'}\,\\
                &\times \Big( e^{-i\,\varepsilon^{\lambda}_{\vec k\vec k'}\,t} 
-1 +i\,\varepsilon^{\lambda}_{\vec k\vec k'}\,t 
\Big)\Big(\varepsilon^{\lambda}_{\vec k\vec k'}\Big)^{-2}  \, .
\end{split}
        \label{eq:L2}
\end{align}
Initially it decays quadratically with time, as obvious from the double 
integral from $0$ to $t$ in Eq.\eqref{eq:diag_uneval}, and also reflected in 
Eq.\eqref{eq:L2} by the subtraction of the first two terms in the expansion of 
the exponential. 

More relevant is the long-time behavior of the real part     
 \begin{equation}
      \text{Re} L_2(t) = -\sum_{\lambda,\vec k,\vec k'} \Big| W^{\lambda}_{\vec 
      k,\vec k'}\Big|^2 (1-f^{\lambda}_{\vec k})f^{\lambda}_{\vec k'}\,
                \,\frac{1-\cos(\varepsilon^\lambda_{\vec k 
\vec k'}\, t)}{(\varepsilon^\lambda_{\vec k \vec k'})^2} \, ,
        \label{eq:real_L2}
\end{equation}
which controls the polarization decay. Indeed, using the large $t$ asymptotics 
of $(1-\cos\omega t)/\omega^2 \sim \pi\, \delta(\omega)\, t$, one finds an 
exponential attenuation $\mathcal P(t) \sim \exp(-\Gamma t)$ with the decay 
rate given by
\begin{equation}
\Gamma = \pi \sum_{\lambda,\vec k,\vec k'} \Big| W^{\lambda}_{\vec 
      k,\vec k'}\Big|^2 (1-f^{\lambda}_{\vec k})f^{\lambda}_{\vec 
k'}\,\delta(\varepsilon^\lambda_{\vec k \vec k'})\, .
      \label{eq:Gamma}
\end{equation}

A comparative discussion with the IBM is in order. The dephasing process does 
not imply a change of population (pure dephasing) and therefore the decay rate 
$\Gamma$ does not involve energy transfer, as seen by the presence of the 
$\delta$-function. In the case of IBM that means only zero-energy phonons are 
involved. Then all the discussion takes place around the spectral edge, and 
depends 
on the density of states and coupling constants there. Usually they vanish as 
a higher power of energy and overcome the singularity of the Bose-Einstein 
distribution, with the result that $\Gamma=0$. This leads to the problem of 
the zero-phonon line (ZPL) appearing as an artificial pure $\delta$-peak in the 
spectrum. This is a weak point and several ways out have been devised, like 
including a phenomenological line broadening \cite{zimm_ebg}, a phonon-phonon 
interaction \cite{zimmulj}, or considering a lower dimensionality \cite{wacker} 
to enhance the density of states. The fermionic case is free from this problem, 
since $\Gamma$ relies on quantities around the chemical potential.

Also, it is worth noting that limiting the expansion to $L_2$ gives the exact 
result in the IBM, while here it is only an approximation. One may plead in 
favor of neglecting higher diagrams by arguing that a lot of compensation takes 
place between the direct terms in Eq.\eqref{eq:W} and the exchange terms are 
small, in other words the QD-WL coupling is weak. Nevertheless, this is not 
sufficient, since it might turn out that higher order diagrams behave as 
higher powers in time, and thus asymptotically overtake the second order one.
We argue below that this is not the case.

Indeed, the structure of the diagrams is such that the $\theta$-functions 
contained in the Green's functions splits the expression into integrals of the 
form
\begin{equation}
I_n(t) = \int_0^t \text{d}t_1 e^{i \omega_1 t_1}\int_0^{t_1} \text{d}t_2 e^{i 
\omega_2 t_2} \,\ldots \int_0^{t_{n-1}} \text{d}t_n e^{i \omega_n t_n} \, .
\label{eq:In}
\end{equation}
The frequency appearing in each time integration is the difference of the 
energies of the Green's functions meeting at the corresponding internal points. 
Summing these pairwise differences around the diagram entails the relation 
$\omega_1+\omega_2+ \ldots + \omega_n=0$.    

On the other hand, the Laplace transform of $I_n(t)$ can be easily calculated 
and gives
\begin{align}
J_n(s) = &\frac{1}{s} \, 
\frac{1}{s+i\omega_1}\,\frac{1}{s+i(\omega_1+\omega_2)} 
          \cdots \nonumber \\
         &\cdots \frac{1}{s+i(\omega_1+\omega_2+ \ldots +\omega_n)}
         \label{eq:Jn}
\end{align}
The last factor is actually $1/s$, like the first, so that $J_n(s) \sim 1/s^2$ 
around $s=0$. This corresponds to a behavior $I_n(t) \sim t$ as $t \to \infty$ 
for all $n$. For instance, the $n=2$ case, discussed above, can be recovered 
from $J_2(s) = 1/s^2 \,\cdot 1/(s+i \omega_1)$. The low-$s$ asymptotics of its 
real part generates the linear large time  behavior times the 
$\pi \delta(\omega_1)$ factor, with $\omega_1= \varepsilon^\lambda_{\vec k, 
\vec k'}$.   

Using $g$ as a generic notation for the coupling strength, we conclude that the 
contribution of the diagram of order $n$ at large times is $\sim g^n\,t$. This 
is in agreement with the so-called weak interaction limit \cite{spohn}, stating 
that when $g\to 0$ and simultaneously $t \to \infty$ so that $g^2\,t$ remains 
constant, the Born-Markov dissipative evolution becomes exact. Indeed, here all 
$n>2$ contributions vanish in this scaling limit.

\section{Numerical example} 

As an illustration we consider an InAs/GaAs heterostructure, with a 
self-assembled QD on a WL of $L=2.2$ nm width.
The relevant material parameters are those of Vurgaftman {\em et al.} 
\cite{vurgaftman}. We assume the wavefunctions to be factorized into the 
square-well solution $\chi^\lambda(z)$ along the growth direction
and the in-plane function. The latter are taken as oscillator ground-state 
gaussians for the QD $s$-states for electrons and holes, and as plane waves, 
orthogonalized on the former, for the WL extended states. The gaussians are 
defined by their width $\alpha_\lambda$ in the reciprocal space, i.e. 
$\varphi^\lambda_s(\vec r)= \alpha_\lambda/\sqrt{\pi} \exp(-\alpha^2_\lambda 
r^2/2) $ with $\vec r$ here the in-plane position.
These parameters depend on many geometric and composition features of the QD, 
so that they can reach a broad set of values. For the sake of our example we 
take $\alpha_e=0.2/\text{nm}$ and $\alpha_h=0.1/\text{nm}$.

The phonon-induced dephasing is expected to be less important at low 
temperatures. The Coulomb-assisted dephasing depends on both temperature and 
WL-carrier concentration, therefore lowering the temperature and increasing the 
concentration it has a chance to compete with the phononic processes. 
We consider only the neutral charging, with electrons and holes in the 
WL at the same concentration $n$.

\begin{figure}[t]%
	\includegraphics*[width=\linewidth,height=0.8\linewidth]{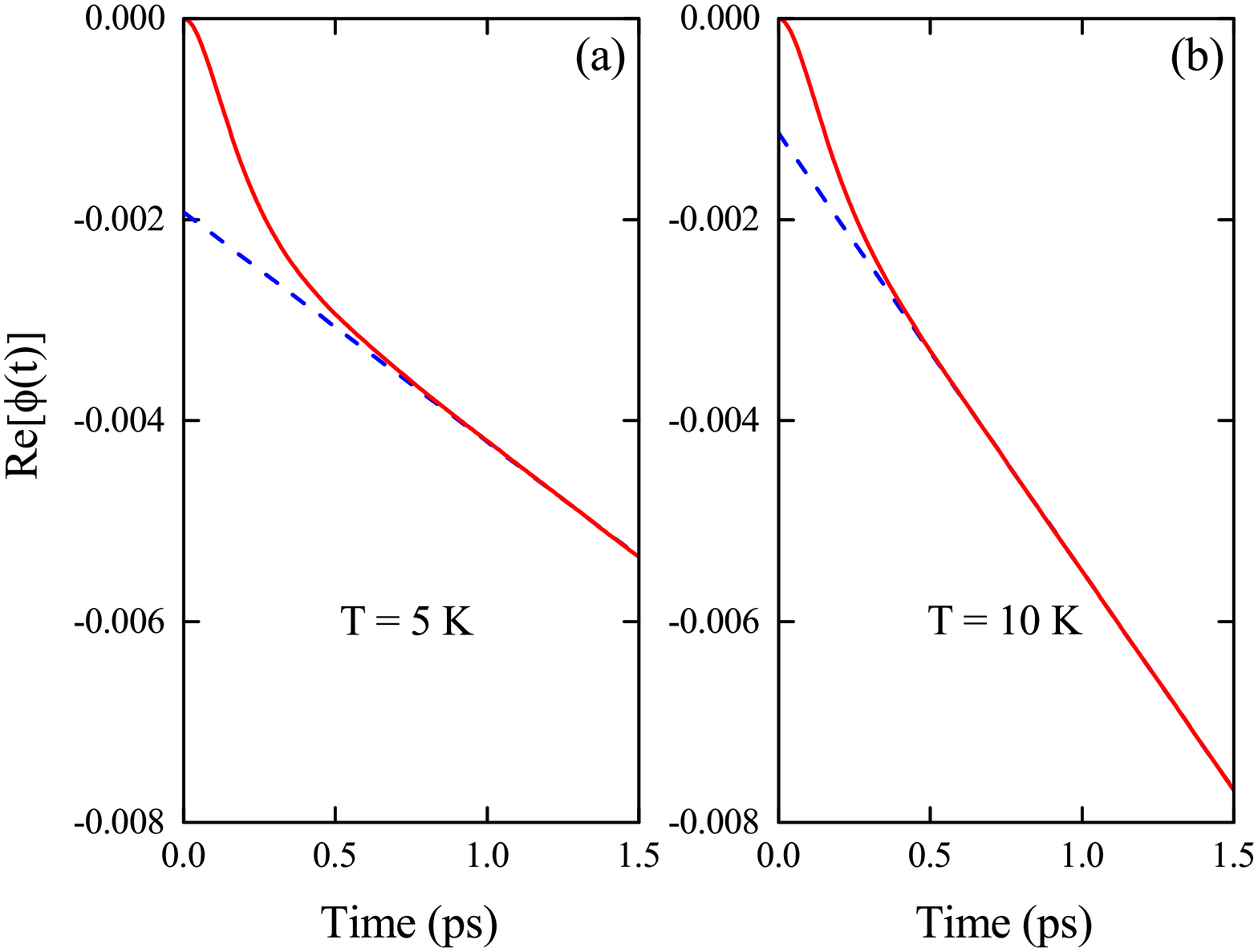}
	\caption{(color online) Time evolution of the real part of $\Phi$ for two 
temperatures,
	 5K (a) and 10K (b) at the same carrier concentration 
$n=10^{12}/\text{cm}^2$.
	 The dephasing reaches a linear decay whose rate increases with 
         temperature.}
	\label{fig:fig2}
\end{figure}

\begin{figure}[t]%
	\includegraphics*[width=\linewidth,height=0.8\linewidth]{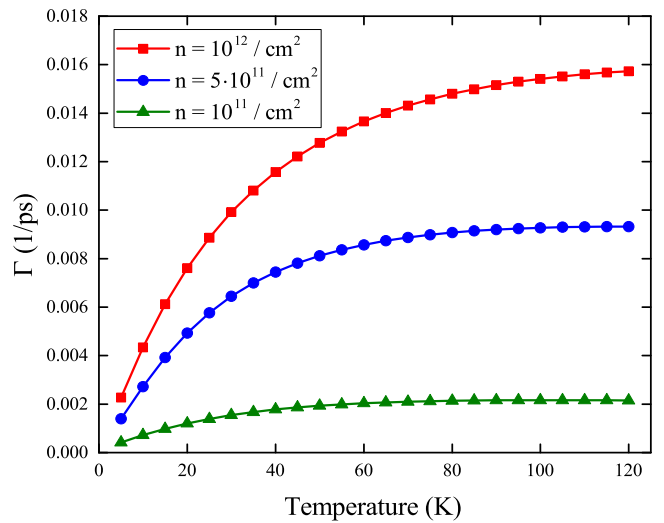}
	\caption{(color online) Dephasing rates $\Gamma$ as functions of 
         temperature for different carrier concentrations.}
	\label{fig:fig3}
\end{figure}

In Fig.~\ref{fig:fig2} the time evolution of the real part of $\Phi(t)$ is 
plotted for two 
different temperatures. The inital quadratic behavior is followed by a linear 
decrease, whose slope $\Gamma$ is the dephasing rate predicted by  
Eq.\eqref{eq:Gamma}. It increases with temperature, as confirmed by 
Fig.~\ref{fig:fig3}, which  shows the values of $\Gamma$ at various 
temperatures and carrier concentrations. 

The range of those values is such that $\hbar \Gamma$ is of the 
order of a few $\mu \text{eV}$. This is comparable with results for dephasing 
by phonons at low temperatures both in theoretical simulations \cite{zimm_ebg, 
takagahara} and in experimental data \cite{woggon, borri_prl}. Experimental 
data obtained by four-wave mixing \cite{borri_prl} do not separate phonon and 
injected carrier contributions to dephasing, but their total effect is still in 
the $\mu \text{eV}$ range.  

For an increase of temperatures from 5K to 120K the dephasing grows by roughly 
one order of magnitude. In the same conditions the rate of dephasing by phonons 
gains two orders of magnitude \cite{zimm_ebg, woggon}, showing a higher 
sensitivity to temperature. Yet in the case of the fermionic bath the decay is 
enhanced also by increasing a second controllable parameter, the carrier 
concentration. 

As mentioned above, the description of the phonon dephasing by the IBM runs 
into the ZPL problem. As seen in Refs. \cite{li_jqe,zimmulj} the slope of the 
long-time 
linear asymptotics is zero, for reasons discussed Sec.\ref{sec:model}. This 
leads to an unphysical pure $\delta$-spike in the frequency domain.

\begin{figure}[t]%
	\includegraphics*[width=\linewidth,height=0.8\linewidth]{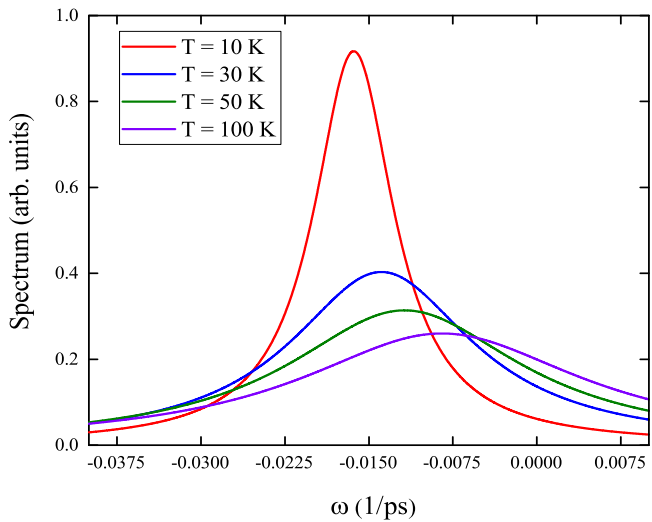}
	\caption{(color online) Absorption spectra for $n=10^{12}/\text{cm}^2$ 
         at T=10, 30, 50, 100K}
	\label{fig:fig4}
\end{figure}

This is not the case with the fermionic bath, as also seen in 
Fig.~\ref{fig:fig4}. 
The main feature of the spectra is their Lorentzian shape, as a consequence of 
the exponential decay in the time domain. Still, the quadratic initial behavior 
replaces the cusp at $t=0$ of a pure exponential by a smooth matching. In the 
frequency domain this leads to a departure from the Lorentzian, in the sense of
a faster decay at large frequencies.

\begin{figure}[t]%
	\includegraphics*[width=\linewidth,height=0.8\linewidth]{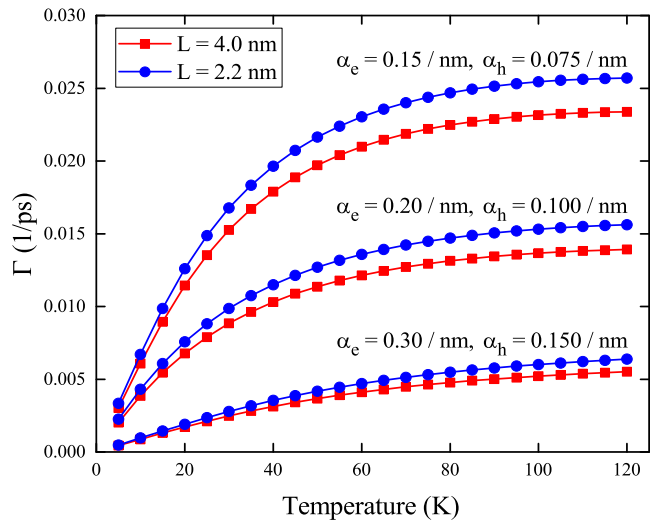}
	\caption{(color online) Temperature dependence of the dephasing rate for 
different WL width $L$ and space extension parameters $\alpha$. All curves for 
$n=10^{12}/\text{cm}^2$}
	\label{fig:fig5}
\end{figure}

In Fig.~\ref{fig:fig5} we also consider the dependence of dephasing on other, 
more geometric parameters. It is seen that $\Gamma$ is not very sensitive to
the WL width $L$, but it is significantly influenced by the spatial extension 
of the QD $s$-states. The broader the states, the stronger the dephasing, due 
to a more efficient scattering.

\section{Conclusions}
 
 In conclusion, we have shown that a fermionic counterpart of the popular IBM 
is possible. It describes the QD exciton interaction with the fermionic bath 
consisting of injected carriers in the bulk or WL. 
Similarities and differences to the IBM are pointed out. For instance, the 
present solution takes the form of a diagrammatic series expansion, while the 
IBM is exact, but this advantage is lost as soon as other interactions are 
present. Also, our case is free from the ZPL problem inherent to the bosonic 
case.
The dephasing process is controlled not only by temperature but also by the 
chemical potential of the bath. The numerical illustration shows that at low 
temperatures and higher carrier concentrations
the dephasing times are comparable with those produced by the phonon 
interaction. But, of course, this is also dependent on the parameters of the 
particular case considered. The dephasing gets stronger at higher temperature 
and concentration, as well as with broader charge distribution of QD states.

\section*{Acknowledgments}
The authors acknowledge financial support from CNCS-UEFISCDI Grant No. 
PN-III-P4-ID-PCE-2016-0221
(I.V.D. and P.G.) and from the Romanian Core Program PN19-03, Contract No. 21
N/08.02.2019, (I.V.D. and M. \c T.).

   \appendix
   \bibliography{mss}

\end{document}